\documentclass[11pt,twoside]{article}
\usepackage[utf8]{inputenc}
\usepackage{times,fancyhdr}
\usepackage[dvips]{graphicx}
\usepackage{latexsym}
\usepackage[affil-it]{authblk}
\usepackage{amsmath}
\usepackage{amssymb}
\usepackage{mathtools}
\usepackage{color}
\usepackage[title, titletoc]{appendix}

\usepackage{hyperref}

\usepackage[backend=biber,style=phys,biblabel=brackets,pageranges=false,eprint=true]{biblatex}
\DeclareSourcemap{
 \maps[datatype=bibtex,overwrite=true]{
  \map{
    \step[fieldsource=Collaboration, final=true]
    \step[fieldset=usera, origfieldval, final=true]
  }
 }
}

\renewbibmacro*{author}{%
  \printnames{author}
  \iffieldundef{usera}{
  }{
    (\printfield{usera} Collaboration)
  }
}

\DeclareFieldFormat{eprint:arxiv}{%
    arXiv\addcolon
    \ifhyperref
    {\href{https://arxiv.org/abs/#1}{%
        #1%
    }}
    {%
        #1%
    }}

\usepackage[top=4cm, bottom=4cm, left=4cm, right=4cm]{geometry}

\begin{document}

\title{The role of the chemical potential in coupling superfluid dark matter to baryons}
\author{Tobias Mistele}
\affil{\small Frankfurt Institute for Advanced Studies\\
Ruth-Moufang-Str. 1,
D-60438 Frankfurt am Main, Germany
}
\date{}
\maketitle

\begin{abstract}
Superfluid dark matter postulates that the centers of galaxies contain superfluid condensates.
An important quantity regarding these superfluids is their chemical potential $ \mu $.
Here, we discuss two issues related to this chemical potential.
First, there is no exactly conserved quantity associated with this chemical potential due to the symmetry-breaking baryon-phonon coupling.
Second, $ \mu $ is sometimes introduced by shifting the phonon field by $ \mu \cdot t $ which -- again due to the symmetry-breaking baryon-phonon coupling -- introduces an explicit time dependence in the Lagrangian.
We investigate under which conditions introducing a chemical potential is nevertheless justified and show how to correctly introduce it when these conditions are met.
We further propose a model that recovers superfluid dark matter's zero-temperature equations of motion including a chemical potential even if the aforementioned conditions for justifying a chemical potential are not met.
\end{abstract}

\section{Introduction}
\label{sec:introduction}

Superfluid dark matter  ({\sc SFDM}) has recently been proposed as an explanation for the missing non-baryonic mass on both galactic and cosmological scales \cite{Berezhiani2015, Berezhiani2018}.
The main idea is that the dark matter particles condense to a superfluid on galactic scales, where phonons then exert an additional force on the baryons.
While superfluid condensates of light dark matter particles have been considered before on general grounds \cite{Sikivie2009,Noumi2014,Davidson2013,DeVega2014,Davidson2015,Guth2015,Aguirre2016,Dev2017,Eby2018,Sarkar2018}, the additional phonon-mediated force is specific to the type of superfluid proposed by \cite{Berezhiani2015, Berezhiani2018}.
This additional force is similar to the gravitational force of Modified Newtonian Dynamics ({\sc MOND}) \cite{Milgrom1983a,Milgrom1983c,Bekenstein1984}.
As a result, {\sc SFDM} can reproduce the successes of {\sc MOND} on galactic scales.
On cosmological scales, the superfluid breaks down such that it behaves like ordinary cold dark matter (CDM) and can reproduce the successes of $ \Lambda $CDM on cosmological scales.

The {\sc MOND}-like force on galactic scales requires a direct coupling between phonons and baryons (but not between phonons and photons \cite{Boran2018, Sanders2018,Hossenfelder2019}).
This coupling breaks the $ U(1) $ symmetry usually associated with superfluidity \cite{Schmitt2015}.
This causes two problems regarding {\sc SFDM}'s chemical potential.

The first problem is that {\sc SFDM} might not have a chemical potential at all.
This is because the $ U(1) $ symmetry is broken in {\sc SFDM} and chemical potentials as a statistical physics concept are a consequence of conserved quantities.
This problem requires a solution, since {\sc SFDM} needs a chemical potential for its phenomenology on galactic scales.

The other problem is that {\sc SFDM}'s chemical potential $ \mu $ is sometimes introduced by shifting the phonon field by $ \mu \cdot t $ (see e.g. Ref.~\cite{Berezhiani2015} below Eq.~(6) for an explicit example).
This leads to an explicit time dependence in the $ U(1) $-breaking baryon-phonon coupling.
Phenomenologically, this may or may not be problematic depending on the size of $ \mu $ and the details of the model.
But conceptually, a chemical potential is an equilibrium quantity from statistical physics and should not be associated with an explicit time dependence.
This problem requires an explanation.

The aim of this paper is to clarify these issues regarding {\sc SFDM}'s chemical potential.
In the following, we employ units with $ c = \hbar = 1 $ and the metric signature $ (+, -, -, -) $.
Small Greek indices run from $ 0 $ to $ 3 $ and denote spacetime dimensions.

We start with an introduction to {\sc SFDM} in Sec.~\ref{sec:setup}.
In Sec.~\ref{sec:nonconservation}, we discuss the nonconservation of the $ U(1) $ charge of {\sc SFDM} and the conditions under which a chemical potential may nevertheless be introduced.
We then show how to correctly introduce a chemical potential in case these conditions are met in Sec.~\ref{sec:chempot}.
Using these results, we distinguish two different non-relativistic limits which may be taken in {\sc SFDM} in Sec.~\ref{sec:nonrel}.
In Sec.~\ref{sec:emt}, we address possible confusions regarding {\sc SFDM}'s equilibrium energy-momentum tensor.
Finally, we propose an alternative model which avoids the problems of {\sc SFDM} regarding its chemical potential in Sec.~\ref{sec:alternative}.
We conclude in Sec.~\ref{sec:conclusion}.

\section{The model of {\sc SFDM}}
\label{sec:setup}

In this section, we will introduce the formalism behind {\sc SFDM}.
To this end, we will first discuss an illustrative toy model and then introduce a more general class of models which we will consider in the rest of this paper.

{\sc SFDM} models are usually based on a complex scalar field $ \phi $ \cite{Berezhiani2015, Berezhiani2018},
\begin{align}
 \phi = \frac{\rho}{\sqrt{2}} \exp(-i \theta) \,.
\end{align}
As a simple example, consider the Lagrangian \cite{Berezhiani2015}
\begin{align}
 \label{eq:Lex}
 \mathcal{L}_{\textrm{ex}} = K - \frac12 m^2 \rho^2 - \frac16 \lambda_6 \rho^6 - \lambda \, \theta \, \rho_b \,,
\end{align}
with the kinetic term $ K $,
\begin{align}
 \label{eq:K}
  K = \frac12 \nabla_\alpha \rho \, \nabla^\alpha \rho + \frac12 \rho^2 \,  \nabla_\alpha \theta \, \nabla^\alpha \theta  \,.
\end{align}
Here, $ \rho_b $ is the baryonic density and $ m > 0 $, $ \lambda_6 > 0 $, and $ \lambda $ are constants with mass dimensions $1$, $-2$, and $0$, respectively.
For simplicity, we assume a flat background spacetime.
The baryon-phonon coupling $ -\lambda \, \theta \, \rho_b $ is the only term in $ \mathcal{L}_{\textrm{ex}} $ which breaks the $ U(1) $ symmetry $ \theta \to \theta + \textrm{const.} $
This will be discussed in detail below.
We do not include the dynamics of the baryons since they are not important for our arguments.

The core idea of {\sc SFDM} is that the $ \phi $ particles condense to a superfluid in the centers of galaxies.
This condensation is induced by a large enough chemical potential $ \mu $.
Here, we introduce this chemical potential by shifting $ \theta \to \theta + \mu \cdot t $.
This procedure of introducing a chemical potential is actually incorrect as discussed below in Sec.~\ref{sec:chempot}.
We nevertheless follow this procedure for now to illustrate its problems regarding the baryon-phonon coupling.
We can then see that the effective potential for $ \rho $,
\begin{align}
 V_{\textrm{eff}}(\rho) = \frac12 \rho^2 \left(m^2 + (\vec{\nabla} \theta)^2 - (\dot{\theta} +\mu)^2 \right) + \frac16 \lambda_6 \rho^6 \,,
\end{align}
has its minimum at nonzero $ \rho $ for $ \mu > m $ and if derivatives of $ \theta $ are small enough.
As long as derivatives of $ \rho $ are negligible, this minimum of $ V_{\textrm{eff}}(\rho) $ approximately solves the equation of motion for $ \rho $ and describes the superfluid condensate.
It is given by
\begin{align}
 \label{eq:rhomin}
 \sqrt{\lambda_6} \, \rho^2 = \sqrt{(\dot{\theta} + \mu)^2 - (\vec{\nabla} \theta)^2 - m^2} \,.
\end{align}
We can now see the connection between the superfluid phase of {\sc SFDM} and {\sc MOND}.
Neglecting derivatives of $ \rho $, the effective Lagrangian for $ \theta $ becomes
\begin{align}
 \label{eq:thetaL}
 \mathcal{L}_{\textrm{eff},\theta} = \frac13 \frac{1}{\sqrt{\lambda_6}} \left( (\dot{\theta} +\mu)^2 - (\vec{\nabla} \theta)^2 - m^2\right)^{3/2} - \lambda \, (\theta + \mu \cdot t) \, \rho_b \,,
\end{align}
which has the characteristic {\sc MOND}-like power of 3/2 in the kinetic term \cite{Bekenstein1984}.\footnote{
This possibility to obtain {\sc MOND}-like behavior from a standard complex scalar field was first noted in the context of phase coupling gravitation \cite{Bekenstein1988a, Bekenstein1988b}.}
Unfortunately, this model does not give the usual {\sc MOND} phenomenology since spatial gradients cannot dominate in $ \mathcal{L}_{\textrm{eff},\theta} $.
Therefore, it is not a realistic model of {\sc SFDM}, as discussed in Ref.~\cite{Berezhiani2015}.
Still, it serves to introduce the ideas behind {\sc SFDM}.

As mentioned above, $ \mathcal{L}_{\textrm{eff}, \theta} $ has an explicit time dependence $ \mu \cdot t $.
This is the result of introducing the chemical potential $ \mu $ by shifting $ \theta \to \theta + \mu \cdot t $ in the symmetry-breaking baryon-phonon coupling $ - \lambda \, \theta \, \rho_b $.
Below, we will argue that this time dependence is an artifact of incorrectly introducing the chemical potential.
Conceptually, this is important because equilibrium quantities should not be associated with an explicit time dependence.
Phenomenologically, the term $ - \lambda \, \mu \, t \, \rho_b $ may also be problematic depending on the size of $ \lambda \mu $ and the physical situation under consideration.
In particular, we will see below that the timescale $ (\lambda \mu)^{-1} $ of this time dependence may be significant for galaxies.

In the following, we consider Lagrangians of the form
\begin{align}
 \label{eq:L}
 \mathcal{L}(\dot{\theta}, \vec{\nabla} \theta, \theta, \dot{\rho}, \vec{\nabla} \rho, \rho) &= f(K, \rho) - \lambda \, \theta \, \rho_b \,,
\end{align}
where $ f $ is some function of $ K $ and $ \rho $.
This allows for Lagrangians with both standard and non-standard kinetic terms.
Examples of both cases are discussed in Ref.~\cite{Berezhiani2018}.
In Ref.~\cite{Berezhiani2018}, $ \lambda $ is parametrized as $ \alpha \, \Lambda / M_{\textrm{Pl}} $ where $ M_{\textrm{Pl}} $ is the Planck mass, $ \alpha $ is a dimensionless coupling constant, and $ \Lambda $ is related to the self-interaction strength (for the toy model from Eq.~\eqref{eq:Lex}, $ \Lambda = 1/(2 \sqrt{\lambda_6}) $).
For our numerical estimates, we will use the fiducial numerical values from Ref.~\cite{Berezhiani2018}, which give $ \lambda \approx 10^{-31} $.
We will often suppress arguments of $ \mathcal{L} $ and use the short-hand $ \mathcal{L}(\dot{\theta}, \theta) $ when only the dependence on $ \dot{\theta} $ and $ \theta $ is relevant for us.

What we refer to as the chemical potential $ \mu $ is the relativistic chemical potential.
This differs from the non-relativistic chemical potential $ \mu_{\textrm{non-rel}} $ by the mass $ m $ of the particles of which the superfluid consists \cite{Schmitt2015}.
For $ \mu > 0 $,
\begin{align}
 \mu = m + \mu_{\textrm{non-rel}} \,.
\end{align}
This is in contrast to Refs.~\cite{Berezhiani2015, Berezhiani2018} where $ \mu $ denotes the non-relativistic chemical potential.
Here, we take $ \mu $ to be the relativistic chemical potential to avoid any confusion between introducing a chemical potential $ \mu = m + \mu_{\textrm{non-rel}} $ with $ |\mu_{\textrm{non-rel}}| \ll m $ and considering non-relativistic particle-like solutions in vacuum with $ \theta = m \cdot t + \theta_{\textrm{non-rel}} $ and $ |\dot{\theta}_{\textrm{non-rel}}| \ll m $.
These two non-relativistic limits are closely related without the baryon-phonon coupling, i.e. with $ \lambda = 0 $.
However, with nonzero baryon-phonon coupling one needs to be careful.
This will be further discussed in Sec.~\ref{sec:nonrel}.

\section{Nonconservation of charge}
\label{sec:nonconservation}

We will now discuss the first main point of this paper, namely under which conditions introducing a chemical potential is justified in {\sc SFDM}.
As mentioned above, the Lagrangian $ \mathcal{L} $ from Eq.~\eqref{eq:L} has a shift symmetry $ \theta \to \theta + \textrm{const.} $, if we set the baryon-phonon-coupling to zero, i.e. if $ \lambda = 0 $.
In this case, there is a corresponding conserved current $ j^\alpha $,
\begin{align}
 \label{eq:j}
 j^\alpha = \frac{\partial \mathcal{L}}{\partial (\nabla_\alpha \theta)} = \frac{\partial f}{\partial K} \, \rho^2 \, \nabla^\alpha \theta \,,
\end{align}
whose conservation is equivalent to the equation of motion of $ \theta $.
Such conservation laws restrict the accessible phase space of a system and, therefore, treating such a system with statistical physics methods requires a corresponding chemical potential.

However, in {\sc SFDM}, $ \lambda $ is nonzero so that we have $ \nabla_\alpha j^\alpha \neq 0 $.
Explicitly,
\begin{align}
 \label{eq:thetaeom}
 \nabla_\alpha j^\alpha = - \lambda \rho_b \,.
\end{align}
Assuming $ \vec{j} $ falls off fast enough at spatial infinity, this gives
\begin{align}
 \label{eq:dotQ}
 \dot{Q} = - \lambda \int d^3\vec{x} \, \sqrt{-g} \, \rho_b \equiv - \lambda M_b
\end{align}
for the time dependence of the total charge $ Q $.
Here, $ g $ is the metric determinant, $ Q \equiv \int d^3\vec{x} \, \sqrt{-g} \, j^0 $, and $ M_b $ is the total baryonic mass.
We can use this result to get an estimate for $ | \dot{Q}/Q | $ on galactic scales.
To this end, we take the non-relativistic limit and estimate $ |Q| \approx M_{\textrm{DM}}/m $.\footnote{
    This overestimates $ Q $ if there are $ \phi $ particle-antiparticle pairs.
    In this case, the actual $ |\dot{Q}/Q| $ is even larger which is even more problematic for {\sc SFDM}.
    However, for an equilibrium superfluid with a chemical potential $ \mu > m $ there are much more particles than antiparticles.
}
This gives
\begin{align}
 \label{eq:tQ}
 \left| \frac{\dot{Q}}{Q} \right| \approx \lambda \, m \frac{M_b}{M_{\textrm{DM}}} = \frac{1}{t_Q} \approx \frac{1}{10^8\,\textrm{yr}} \frac{M_b}{M_{\textrm{DM}}} \,,
\end{align}
with the definition $ t_Q^{-1} \equiv \lambda \, m \, (M_b/M_{\textrm{DM}}) $ and the fiducial numerical values from Ref.~\cite{Berezhiani2018}, i.e. $ \lambda \approx 10^{-31} $ and $ m = 1\,\textrm{eV} $.
This means that, on timescales much shorter than $ t_Q \approx (M_{\textrm{DM}}/M_b)\cdot10^8\,\textrm{yr} $, we can take $ Q $ to be approximately conserved.
Consequently, it may be justified to consider an approximate equilibrium -- valid for times much shorter than $ t_Q $ -- that assumes conservation of $ Q $ and includes a corresponding chemical potential.
For large times, this approximate equilibrium breaks down and the system may reach the perfect equilibrium that does not assume conservation of $ Q $ and does not include a chemical potential associated with $ Q $.
Whether the approximate or the perfect equilibrium is relevant depends on the physical situation under consideration.

In the case of galactic rotation curves, the timescale $ t_Q $ should be compared to the dynamical time of galaxies,
\begin{align}
 t_{\textrm{dyn}} = \frac{1}{\sqrt{G \cdot \rho_{\rm{tot}}}} \,,
\end{align}
where $ G $ is the gravitational constant and $ \rho_{\rm{tot}} $ is the total energy density.
For galaxies that satisfy $ t_Q \gg t_{\textrm{dyn}} $, we can neglect the nonconservation of $ Q $, introduce a chemical potential, and find the usual {\sc SFDM} phenomenology of galactic rotation curves.
However, a typical order of magnitude of $ t_{\textrm{dyn}} $ is $ 10^8 \, \textrm{yr} $, which is not necessarily much smaller than $ t_Q \approx (M_{\textrm{DM}}/M_b) \cdot10^8\,\textrm{yr} $.
Therefore, the nonconservation of $ Q $ can be significant on timescales relevant for galactic rotation curves.
In this case, the concept of a chemical potential may not be applicable such that the usual {\sc SFDM} phenomenology of galactic rotation curves does not follow.

Note that our estimate $ |\dot{Q}/Q| \approx t_Q^{-1} $ concerns galaxies as a whole.
In principle, parts of a galaxy could stay in the approximate equilibrium much longer than $ t_Q $.
However, as we will discuss in Sec.~\ref{sec:emt}, there is a local analogue $ t_{\rm{loc}} = (\lambda \, m \, (\rho_b/\rho_{\rm{tot}}))^{-1} $ of $ t_Q $.
In the superfluid cores of galaxies, $ t_{\rm{loc}} $ is typically even smaller than $ t_Q $.
Therefore, we do not expect the approximate equilibrium there to be valid much longer than $ t_Q $.\footnote{
    What this argument does not rule out are spherical shells of long-lived approximate equilibrium at intermediate radii.
    This would be a significant deviation from standard {\sc SFDM} phenomenology.
    Studying this possibility is left for future work.
}
In the following, we are not careful to distinguish between $ t_Q $ and $ t_{\rm{loc}} $.
Our main point is that {\sc SFDM}'s approximate equilibrium is not valid indefinitely.
The numerical difference between $ t_Q $ and $ t_{\rm{loc}} $ is secondary for our purposes.

The above argument to allow for a chemical potential on timescales much shorter than $ t_Q $ is similar to the argument why black-body radiation does not rule out massive photons \cite{Bass1955, Torres-Hernandez1985}.
Namely, introducing a nonzero photon mass leads to an additional degree of freedom -- the longitudinal photon.
This is true for arbitrarily small photon masses.
Therefore, even the tiniest photon mass produces an extra factor of $ 3/2 $ in the Stefan-Boltzmann law for a perfect equilibrium.
Since we do not observe this factor of $ 3/2 $ we might then conclude that photons are exactly massless.
However, this would be too quick.
The reason is that the interactions of the longitudinal photon with ordinary matter tend to zero if the photon mass tends to zero.
Therefore, the longitudinal photons cannot equilibrate on short enough timescales.
In particular, for photons in a cavity, the longitudinal polarization is negligible on timescales much shorter than
\begin{align}
 t_{\gamma} = V^{1/3} (E_\gamma / m_\gamma)^2 \,,
\end{align}
where $ V $ is the volume of the cavity, $ E_\gamma $ is the energy of the photons, and $ m_\gamma $ is the photon mass \cite{Bass1955}.
The experimental bounds on $ m_\gamma $ are such that the timescale $ t_\gamma $ is larger than the age of the universe for typical experimental settings.
Therefore, experiments never see a perfect equilibrium involving all three photon polarizations.
Still, it is justified to consider an approximate equilibrium on timescales much shorter than $ t_\gamma $ that neglects the longitudinal photons.
This is why massive photons are not ruled out by black-body radiation measurements.
Here, the timescale $ t_\gamma $ is analogous to the timescale $ t_Q $ introduced above.
In both cases, there is an approximate equilibrium on short enough timescales that has special properties compared to the perfect equilibrium that is reached for infinite times.
In the case of massive photons, the third photon polarization can be neglected in the approximate equilibrium, while in the case of {\sc SFDM} the nonconservation of $ Q $ can be neglected in the approximate equilibrium.

A possible objection to our estimate for $ |\dot{Q}/Q| $ is the following:
Assuming spherical symmetry and a flat background spacetime, there is an exact static solution of Eq.~\eqref{eq:thetaeom},
\begin{align}
 \label{eq:jr}
 j^r(r) = -\frac{\lambda M_b(r)}{4\pi r^2} \,, \quad \dot{Q} = 0 \,.
\end{align}
This is the solution usually assumed in the center of galaxies in {\sc SFDM} \cite{Berezhiani2015, Berezhiani2018}.
For this solution, our estimate for $ |\dot{Q}/Q| $ from Eq.~\eqref{eq:tQ} is obviously incorrect.
However, this solution represents a highly idealized situation and we expect that our estimate for $ |\dot{Q}/Q| $ does hold in more realistic situations.
To see why, suppose that the baryonic density $ \rho_b $ vanishes outside some spatial volume $ V $ and split up the total charge $ Q $ into the charge $ Q_{\textrm{in}} $ inside $ V $ and the charge $ Q_{\textrm{out}} $ outside $ V $.
Eq.~\eqref{eq:thetaeom} then gives
\begin{subequations}
\begin{align}
 \dot{Q}_{\textrm{in}} &= - \int_{\partial V} d\vec{S} \cdot \vec{j} - \lambda M_b \,, \\
 \dot{Q}_{\textrm{out}} &= + \int_{\partial V} d\vec{S} \cdot \vec{j} - \int_{\partial \mathbb{R}^3} d\vec{S} \cdot \vec{j} \,.
\end{align}
\end{subequations}
That is, the change in $ Q_{\textrm{in}} $ is the charge created due to the symmetry-breaking baryon coupling minus the charge that crosses the boundary of $ V $.
Likewise, the change in $ Q_{\textrm{out}} $ is the charge that crosses the boundary of $ V $ minus the charge that crosses spatial infinity.

In the case of the static solution from Eq.~\eqref{eq:jr}, $ \dot{Q} $ vanishes because the charge created due to the symmetry-breaking baryon coupling exactly balances the charge crossing spatial infinity.
In fact, the charge created due to the baryon coupling spreads out in such a way that all observables are constant in time.
That is, this solution represents an equilibrium in which there is a perfect balance between all dynamical processes, leading to exactly constant observables.
Mathematically, the reason our estimate for $ |\dot{Q}/Q| $ fails in this case is that there is charge crossing spatial infinity.
This is because the $ \vec{\nabla} \vec{j} $ term in Eq.~\eqref{eq:thetaeom} does not fall off fast enough and cannot be neglected when deriving our estimate for $ \dot{Q} $.
Since $ \vec{j} $ falls off so slowly, one may suspect that this solution has infinite total energy and charge.
Indeed, for the toy model from Eq.~\eqref{eq:Lex}, it can be shown explicitly that Eq.~\eqref{eq:jr} gives infinite total energy.

This suggests that the solution from Eq.~\eqref{eq:jr} represents an idealized and possibly unphysical situation.
For realistic galaxies, we expect that this solution is not valid at larger radii where thermal equilibrium breaks down \cite{Berezhiani2015, Berezhiani2018}.
Therefore, we expect that $ \vec{j} $ falls off fast enough to give zero charge crossing spatial infinity.
This implies $ \dot{Q} = - \lambda M_b $ and we recover our above estimate for $ |\dot{Q}/Q| $.

More generally -- irrespective of the details of the solution from Eq.~\eqref{eq:jr} -- the existence of one solution with $ \dot{Q} = 0 $ does not imply that we can assume $ Q $ to be conserved for our purposes.
This is because we are interested in the chemical potential of {\sc SFDM} which is an equilibrium quantity of statistical physics.
The central assumption of statistical physics is that, given fixed values for all conserved quantities, the ensemble average over all field configurations satisfying the constraints equals the time average for a single such field configuration.
Therefore, it is justified to take $ Q $ to be conserved only if $ \dot{Q} = 0 $ for all solutions.
Our above estimate for $ |\dot{Q}/Q| $ implies that taking $ Q $ to be conserved may be a good approximation on timescales much shorter than $ t_Q $, but not on longer timescales.

In Sec.\ref{sec:alternative}, we will discuss how it is possible to retain the {\sc SFDM} phenomenology of galactic rotation curves even if $ t_Q \ll t_{\textrm{dyn}} $.
But for now we will assume that $ t_Q \gg t_{\textrm{dyn}} $, such that introducing a chemical potential is justified.

\section{Chemical potential with approximately conserved charge}
\label{sec:chempot}

As discussed in the previous section, $ Q $ is approximately conserved on timescales much shorter than $ t_Q $.
On  these timescales, the approximate conservation of $ Q $ effectively restricts the phase space accessible to the superfluid.
In a statistical physics treatment, we should therefore introduce a chemical potential $ \mu $ associated with $ Q $.
This brings us to the second main point of this paper, namely how to correctly introduce a chemical potential for the approximately conserved charge $ Q $.
To this end, we assume a static background metric so that there is a preferred Hamiltonian $ H $ to which we can apply standard statistical physics methods.
For concreteness, we take
\begin{align}
 \label{eq:metric}
 ds^2 = N(\vec{x})^2 \, dt^2 - h_{jk}(\vec{x}) \, dx^j dx^k\,,
\end{align}
where $ N(\vec{x}) $ is some function and $ h_{jk}(\vec{x}) $ is a 3-dimensional metric.
This is always possible for a static metric.

In the grand canonical ensemble, conservation of $ Q $ entails that equilibrium quantities should not be calculated from the Hamiltonian $ H $ but from the effective Hamiltonian $ H_{\textrm{eff}} $,
\begin{align}
 \label{eq:Heff}
 H_{\textrm{eff}} =  H - \mu \, Q\,,
\end{align}
where $ H $ and $ Q $ are functions of the canonical variables and their momenta,
\begin{align}
 H = H(\pi, \theta), \quad Q = Q(\pi, \theta) \,,
\end{align}
with $ \pi \equiv \partial (\sqrt{-g} \,\mathcal{L}) / \partial \dot{\theta} $.
For simplicity, we have suppressed the dependence on $ \rho $ and its canonical momentum.
More concretely, the grand canonical partition function $ Z $ is
\begin{subequations}
\label{eq:Z}
\label{eq:path}
\begin{align}
Z &= \int \mathcal{D}\pi \int \mathcal{D}\theta \, \exp\left[ \int d(it) \int d^3 \vec{x} \left( \pi \, \dot{\theta} - \sqrt{-g} \, \mathcal{H}_{\textrm{eff}}(\pi, \theta)  \right) \right] \\
&\begin{multlined}[11.5cm]
 = \int \mathcal{D}\pi \int \mathcal{D}\theta \, \exp\left[ \int d(it) \int d^3 \vec{x} \right. \\
  \left. \times \left( \pi \, \dot{\theta} - \sqrt{-g} \, \mathcal{H}(\pi, \theta) + \mu \, \sqrt{-g} \, j^0(\pi, \theta)  \right) \right]\,.
\end{multlined}
\end{align}
\end{subequations}
Here, $ \mathcal{H} $ is the Hamiltonian density associated with $ H $ and $ \mathcal{H}_{\textrm{eff}} $ is that associated with $ H_{\textrm{eff}} $.
Specifically, $ \sqrt{-g} \, \mathcal{H} = \pi \, \dot{\theta} - \sqrt{-g} \, \mathcal{L} $.
This path integral uses the ``imaginary time'' variable $ \tau \equiv i t $, but we write it in terms of $ t $ for simplicity.
We have not indicated the boundary conditions of the integrals since they are not important for our arguments.

By integrating out $ \pi $ in this Hamiltonian path integral, one can obtain a Lagrangian path integral with an effective Lagrangian $ \mathcal{L}_{\textrm{eff}}(\dot{\theta}, \theta) $.
Doing so, one sees that the correct effective Lagrangian is the first of the following options, not the second:
\begin{align}
 \label{eq:Leff}
 \mathcal{L}_{\textrm{eff}}(\dot{\theta}, \theta) &= \mathcal{L}(\dot{\theta} + \mu, \theta) &\textrm{(correct)}\,, \\
 \label{eq:Leffwrong}
 \mathcal{L}_{\textrm{eff}}(\dot{\theta}, \theta) &= \mathcal{L}(\dot{\theta} + \mu, \theta + \mu \cdot t) &\textrm{(incorrect)} \,.
\end{align}
In other words: Instead of shifting all occurrences of $ \theta $ by $ \theta \to \theta + \mu \cdot t $, one should only shift time derivatives of $ \theta $, $ \dot{\theta} \to \dot{\theta} + \mu $, while leaving $ \theta $ without time derivatives unchanged.
These two procedures are not equivalent due to the baryon-phonon coupling $ - \lambda \, \theta \, \rho_b $.
Shifting all occurrences of $ \theta $ as in Eq.~\eqref{eq:Leffwrong} leads to a time-dependent factor $ (\theta + \mu \cdot t) $ in the baryon coupling.
In contrast, shifting only time derivatives of $ \theta $ as in Eq.~\eqref{eq:Leff} does not introduce an explicit time dependence in $ \mathcal{L}_{\textrm{eff}} $.
Thus, the explicit time dependence in $ \mathcal{L}_{\textrm{eff},\theta} $ discussed in Sec.~\ref{sec:setup} is indeed an artifact of incorrectly introducing the chemical potential.

Eq.~\eqref{eq:Leff} follows from a saddle-point approximation.
This is exact when $ \mathcal{L} $ is at most quadratic in $ \dot{\theta} $, otherwise it is an approximation.
Here, our point is not to evaluate under which conditions this is a good approximation.
Instead, our point is that, whenever this approximation is valid, Eq.~\eqref{eq:Leff} is correct while Eq.~\eqref{eq:Leffwrong} is correct only without the baryon-phonon coupling.

More concretely, Eq.~\eqref{eq:Leff} can be derived by following the steps from Refs.~\cite{Kapusta1981, Haber1982, Bilic2008}, where the effective Lagrangian is derived in the case with $ \lambda = 0 $, i.e. without the baryon-phonon coupling.
We just have to verify that all steps still apply in the case with nonzero baryon-phonon coupling.
To this end, we start with the Hamiltonian path integral for the grand canonical partition function $ Z $ from Eq.~\eqref{eq:Z}.
In order to obtain the Lagrangian path integral, we have to integrate out $ \pi $.
We will do so using the above-mentioned saddle-point approximation.
That is, we must solve the equation
\begin{align}
 \label{eq:pigeneral}
 \frac{\partial (\sqrt{-g} \, \mathcal{H})}{\partial \pi}(\pi, \theta) = \dot{\theta} + \mu \, \frac{\partial (\sqrt{-g} \, j^0)}{\partial \pi}
\end{align}
for $ \pi = \pi(\dot{\theta}, \theta) $ and put the result back into the exponent of Eq.~\eqref{eq:path}.
Without the chemical potential $ \mu $, this procedure simply gives the Lagrangian path integral with the original Lagrangian $ \mathcal{L}(\dot{\theta}, \theta) $ in the exponent,
\begin{align}
 Z|_{\mu=0} \propto \int \mathcal{D}\theta \, \exp \left[ \int d(it) \int d^3\vec{x} \, \sqrt{-g} \, \mathcal{L}(\dot{\theta}, \theta) \right] \,.
\end{align}
This is because, for $ \mu = 0 $, Eq.~\eqref{eq:pigeneral} is the usual Hamiltonian equation of motion for $ \dot{\theta} $ and $ \pi \dot{\theta} - \sqrt{-g} \, \mathcal{H} $ then gives the original Lagrangian $ \mathcal{L}(\dot{\theta}, \theta) $ including the $ \sqrt{-g} $ prefactor.
Here, we have again suppressed the $ \rho $ dependence since it is inessential to our argument.

With a nonzero chemical potential, Eq.~\eqref{eq:pigeneral} contains the additional term $ \mu \, \tfrac{\partial (\sqrt{-g} \, j^0)}{\partial \pi} $.
For a {\sc SFDM} Lagrangian $ \mathcal{L} $ as in Eq.~\eqref{eq:L}, we have
\begin{align}
 j^0 =  \frac{\partial \mathcal{L}}{\partial \dot{\theta}} = \frac{\pi}{\sqrt{-g}} \,.
\end{align}
Therefore, Eq.~\eqref{eq:pigeneral} becomes
\begin{align}
 \label{eq:piwithmu}
 \frac{\partial (\sqrt{-g} \, \mathcal{H})}{\partial \pi}(\pi, \theta) = \dot{\theta} + \mu \,.
\end{align}
The solutions $ \pi = \pi(\dot{\theta}, \theta) $ of this equation can be expressed using the solutions of the same equation with $ \mu = 0 $ by shifting time derivatives of $ \theta $ while leaving $ \theta $ without time derivatives unchanged,
\begin{align}
 \pi(\dot{\theta}, \theta) = \pi|_{\mu = 0}(\dot{\theta} + \mu, \theta) \,.
\end{align}
From this, we finally obtain for the Lagrangian path integral
\begin{subequations}
\begin{align}
 Z &\propto \int \mathcal{D} \theta \, \exp\left[ \int d(it) \int d^3\vec{x} \, \sqrt{-g} \, \mathcal{L}(\dot{\theta}+\mu, \theta) \right]  \\
   &\equiv \int \mathcal{D} \theta \, \exp\left[ \int d(it) \int d^3\vec{x} \, \sqrt{-g} \, \mathcal{L}_{\textrm{eff}}(\dot{\theta}, \theta) \right] \,,
\end{align}
\end{subequations}
which completes the derivation of Eq.~\eqref{eq:Leff}.

In addition, Eq.~\eqref{eq:Leffwrong} can also be ruled out because it leads to a complex effective Lagrangian $ \mathcal{L}_{\textrm{eff}} $.
This is because the path integral for the partition function $ Z $ uses the ``imaginary time'' variable $ \tau $ such that $ t = -i \tau $ is imaginary.
In terms of $ \tau $, Eq.~\eqref{eq:Leffwrong} reads
\begin{align}
 \mathcal{L}_{\textrm{eff}}(i \partial_\tau \theta, \theta) &= \mathcal{L}(i \partial_\tau \theta + \mu, \theta -i \, \mu \cdot \tau) \,.
\end{align}
Therefore, the baryon-phonon coupling $ - \lambda \rho_b \theta $ obtains an imaginary part in $ \mathcal{L}_{\textrm{eff}} $,
\begin{align}
 - \lambda \rho_b (\theta - i \mu \tau) \,.
\end{align}
As a consequence, equilibrium expectation values of observables are not, in general, real, which rules out this way of introducing the chemical potential.
A formal change of variables $ \theta \to \theta -i \, \mu \cdot \tau $ may still be possible, but then one needs be careful to take into account that $ \theta $ becomes complex.
In any case, the above derivation of Eq.~\eqref{eq:Leff} shows that the $ \mu $ introduced by such a change of variables would not be the usual chemical potential of statistical physics associated with a conserved quantity.

\section{Non-relativistic limit}
\label{sec:nonrel}

Actual calculations in {\sc SFDM} are usually done in the non-relativistic limit.
In Ref.~\cite{Berezhiani2015}, this non-relativistic limit is taken by first shifting $ \theta \to \theta + m \cdot t $, i.e.
\begin{align}
 \label{eq:nonrelparticle}
 \mathcal{L}(\dot{\theta}, \theta) \to \mathcal{L}(\dot{\theta} + m, \theta + m \cdot t) \,,
\end{align}
and then expanding in $ \dot{\theta}/m $ and $ \dot{\rho}/(m\,\rho) $ assuming both are of the same order of smallness.
This procedure is correct, for example, for non-relativistic particle-like solutions of the form $ \phi \approx e^{-i(\omega \cdot t - \vec{k} \cdot \vec{x})} $ for some $ \omega $ and $ \vec{k} $.
And more generally, for explicitly time-dependent solutions with dominant time dependence $ e^{-i m \cdot t} $.
However, it is not correct for the equilibrium superfluid in the center of galaxies and for perturbations on top of this superfluid.
In this case, the non-relativistic limit should be taken by first shifting only time derivatives of $ \theta $ but not terms without time derivatives of $ \theta $,
\begin{align}
 \label{eq:nonrelchempot}
 \mathcal{L}(\dot{\theta}, \theta) \to \mathcal{L}(\dot{\theta} + m, \theta) \,,
\end{align}
and then expanding in time derivatives of $ \theta $ and $ \rho $ as in the previous case.
Without the baryon-phonon coupling, i.e. with $ \lambda = 0 $, Eq.~\eqref{eq:nonrelparticle} and Eq.~\eqref{eq:nonrelchempot} are equivalent, but for nonzero $ \lambda $ they are not.
Shifting all occurrences of $ \theta $ as in Eq.~\eqref{eq:nonrelparticle} leads to an explicit time dependence $ (m \cdot t + \theta) $ in the baryon-phonon coupling.
In contrast, shifting only time derivatives of $ \theta $ as in Eq.~\eqref{eq:nonrelchempot} does not introduce any explicit time dependence.

The reason to use Eq.~\eqref{eq:nonrelparticle} and Eq.~\eqref{eq:nonrelchempot} in different situations is the following:
In the case of Eq.~\eqref{eq:nonrelparticle}, one is looking for explicitly time-dependent solutions with dominant time dependence $ e^{-i m \cdot t} $.
In the non-relativistic limit, one expects $ |\dot{\theta} - m| \ll m $ and $ |\dot{\rho}| \ll m\,\rho $.
This leads to the procedure referred to by Eq.~\eqref{eq:nonrelparticle}.
In contrast, in the case of Eq.~\eqref{eq:nonrelchempot}, one introduces a chemical potential $ \mu = m + \mu_{\textrm{non-rel}} $ for the equilibrium superfluid assuming $ |\mu_{\textrm{non-rel}}| \ll m $.
Further, one assumes that any perturbations on top of the equilibrium satisfy $ |\dot{\theta}| \ll m $ and $ |\dot{\rho}| \ll m\,\rho $.
This leads to the procedure referred to by Eq.~\eqref{eq:nonrelchempot}.

In the literature on {\sc SFDM}, the two non-relativistic limits from Eq.~\eqref{eq:nonrelparticle} and Eq.~\eqref{eq:nonrelchempot} are not distinguished.
Often, the Lagrangian is given only after having already taken the non-relativistic limit.
However, this may lead to confusion because the non-relativistic limit introduces an explicit time dependence in some physical situations (namely when considering particle-like solutions in vacuum as in Eq.~\eqref{eq:nonrelparticle}) but not in others (namely for an equilibrium superfluid as in Eq.~\eqref{eq:nonrelchempot}).
To avoid this confusion, we suggest to always start with the relativistic Lagrangian and to take the non-relativistic limit only after having explicitly specified which physical situation is considered.

\section{Equilibrium energy-momentum tensor}
\label{sec:emt}

In this section, we will discuss possible confusions regarding {\sc SFDM}'s equilibrium energy-momentum tensor ({\sc EMT}).
{\sc SFDM}'s {\sc EMT} is, in general,
\begin{align}
\label{eq:emt}
T_{\alpha \beta} = \frac{\partial f}{\partial K} \left( \nabla_\alpha \rho \, \nabla_\beta \rho + \rho^2 \, \nabla_\alpha \theta \, \nabla_\beta \theta \right) - g_{\alpha \beta} (f - \lambda \, \theta \, \rho_b ) \,.
\end{align}
The equilibrium {\sc EMT} can be calculated by expressing $ T_{\alpha \beta} $ in terms of canonical variables and their momenta, and using the result in a path integral similar to that for the partition function $ Z $ discussed in Sec.~\ref{sec:chempot},\footnote{
    We do not rewrite Eq.~\eqref{eq:emtpathintegral} as a Lagrangian path integral since $ T_{\alpha \beta} $ depends on $ \pi $ and the usual derivation of the Lagrangian path integral assumes that factors like $ T_{\alpha \beta} $ in the integrand are independent of $ \pi $.
    See e.g Chapter~9.3 of Ref.~\cite{Weinberg1995} for a discussion of this.
}
\begin{align}
\label{eq:emtpathintegral}
\int \mathcal{D}\pi \int \mathcal{D}\theta \, T_{\alpha \beta}(\pi, \theta) \, \exp\left[ \int d(it) \int d^3\vec{x} \left(\pi \,\dot{\theta} - \sqrt{-g} \, \mathcal{H}_{\textrm{eff}}(\pi, \theta) \right) \right] \,.
\end{align}
We assume that, in the zero-temperature limit, a simple saddle-point approximation is sufficient.
That is, we assume that we can calculate all zero-temperature quantities by shifting $ \dot{\theta} \to \dot{\theta} + \mu $ and using the equations of motion from $ \mathcal{L}_{\textrm{eff}}(\dot{\theta}, \theta) = \mathcal{L}(\dot{\theta} + \mu, \theta) $.
If we do this, we come across two possibly confusing findings.
First, the zero-temperature equilibrium superfluid has a nonzero momentum density,
\begin{align}
\label{eq:T0j}
T_{0j} = \frac{\partial f}{\partial K} \rho^2 \, \mu \, \partial_j \theta \,, \quad j = 1,2, 3 \,.
\end{align}
Second, this does not satisfy the continuity equation $ \nabla_\alpha T^{\alpha 0} = 0 $,\footnote{
    The {\sc EMT} derived from $ \mathcal{L}_{\textrm{eff}} $ (not $ \mathcal{L} $) using Noether's theorem does satisfy $ \nabla_\alpha T^{\alpha 0} = 0 $.
    However, this {\sc EMT} is not symmetric and is not obtained from Eq.~\eqref{eq:emt} by shifting $ \dot{\theta} \to \dot{\theta} + \mu $.
}
\begin{align}
\label{eq:Enonconserved}
\nabla_\alpha T^{\alpha 0} = \frac{1}{\sqrt{-g}} \partial_j \left( \sqrt{-g} \, T^{0j} \right) + \Gamma^0_{\alpha \beta} T^{\alpha \beta} = - \lambda \rho_b \, \mu \,.
\end{align}
For both Eq.~\eqref{eq:T0j} and Eq.~\eqref{eq:Enonconserved} we used a static background metric and assumed that all fields are time independent in equilibrium, we only shifted $ \dot{\theta} \to \dot{\theta} + \mu $.
Since we started from a theory that is explicitly time translation invariant, we would have expected the right-hand side of Eq.~\eqref{eq:Enonconserved} to give zero.\footnote{
    If the external field $ \rho_b $ is constant in time.
    Similarly, $ \nabla_\alpha T^{\alpha j} $ vanishes only if $ \rho_b $ is spatially constant.
}
But already the physical origin of the momentum density in Eq.~\eqref{eq:T0j} is not clear.

Consider first this momentum density from Eq.~\eqref{eq:T0j}.
In Sec.~\ref{sec:nonconservation}, we argued that the charge $ Q $ is not exactly conserved on timescales $ t_Q $.
Since the Lagrangian $ \mathcal{L} $ from Eq.~\eqref{eq:L} is time translation invariant, energy is conserved in this process.
However, we expect energy to be spatially redistributed.
This is because we expect the nonconservation of $ Q $ to proceed by charged perturbations being created and interacting with each other.
We expect these perturbations to travel in all possible directions.
At the same time -- in order to satisfy energy conservation -- the static, localized background fields deplete.
As a result, charge is created and total energy is spatially redistributed to a less localized configuration.
This is the physical origin of the momentum density from Eq.~\eqref{eq:T0j}.

The spatial redistribution of energy due to the $ Q $-nonconservation also explains how we could end up with an equilibrium {\sc EMT} that violates the continuity equation $ \nabla_\alpha T^{\alpha 0} = 0 $.
Namely, this redistribution of energy is a time-dependent process, but we assumed an exactly time-independent equilibrium with constant $ Q $ when deriving Eq.~\eqref{eq:T0j} and Eq.~\eqref{eq:Enonconserved}.
More formally, our starting point in Sec.~\ref{sec:nonconservation} was the effective Hamiltonian $ H_{\textrm{eff}} = H - \mu \, Q $.
But this effective Hamiltonian is derived in statistical physics assuming that $ Q $ is exactly conserved.
As discussed in Sec.~\ref{sec:nonconservation}, this is a good approximation only on timescales much shorter than $ t_Q $.
Therefore, we are effectively neglecting time derivatives of order $ 1/t_Q $ if we assume a perfect equilibrium and use the standard statistical physics formalism with $ H_{\textrm{eff}} = H - \mu \, Q $.
These time derivatives are what is needed to satisfy the continuity equation $ \nabla_\alpha T^{\alpha 0}  = 0 $.

Thus, {\sc SFDM}'s approximate equilibrium with approximately conserved $ Q $ is indeed valid only on timescales much shorter than $ t_Q $.
If time derivative of order $ 1/t_Q $ are important, different concepts are needed.
One example might be a general relativistic treatment where self-consistency requires an exactly conserved {\sc EMT}.
However, we expect that we can safely neglect time derivatives of order $ 1/t_Q $ for most non-relativistic applications, if we are interested only in timescales much shorter than $ t_Q $.
For example, we expect that it is a good approximation to calculate the superfluid's energy density using $ H_{\textrm{eff}} $ if we are interested in galactic rotation curves with $ t_Q \gg t_{\textrm{dyn}} $.

In the case of black-body radiation of massive photons in a cavity (see Sec.~\ref{sec:nonconservation}), we similarly expect that the continuity equation $ \nabla_\alpha T^{\alpha 0} = 0 $ is not exactly fulfilled.
This is because the usually assumed approximate equilibrium without longitudinal photons neglects the creation of longitudinal photons when transverse photons are reflected at the cavity walls \cite{Bass1955}.
In this case, we effectively neglect time derivatives of order $ 1/t_\gamma $, just as we neglect time derivatives of order $ 1/t_Q $ in {\sc SFDM}.
However, one usually doesn't encounter this peculiarity in the literature on massive photons.
The reason is that most actual calculations of black-body radiation of massive photons assume a free photon gas without interactions, i.e. the interactions with the cavity walls are completely neglected.
Neglecting interactions may be a good approximation for massive photons in a cavity, but it is not useful in {\sc SFDM}, since interactions are essential for {\sc SFDM}'s phenomenology.

To be more precise, the neglected time derivatives in {\sc SFDM} may actually deviate from $ 1/t_Q $.
This is because the timescale $ t_Q $ concerns galaxies as a whole, while the continuity equation $ \nabla_\alpha T^{\alpha 0} = 0 $ is a local equation.
Indeed, consider the following estimate.
The time derivatives needed to satisfy the continuity equation $ \nabla_\alpha T^{\alpha 0} = 0 $ can be time derivatives either of $ T^{00} $ or of the metric.
In the non-relativistic limit, we estimate the corresponding terms in $ \nabla_\alpha T^{\alpha 0} $ as $ \dot{\rho}_{\rm{tot}} $ and $ \rho_{\rm{tot}} \cdot \dot{\phi}_N $, respectively.
Here, $ \phi_N $ is the Newtonian gravitational potential.
Dividing by $ \rho_{\rm{tot}} $, these two terms become $ \dot{\rho}_{\rm{tot}}/\rho_{\rm{tot}} $ and $ \dot{\phi}_N $, respectively.
Together, they have to compensate the $ - \lambda \, \mu \, \rho_b / \rho_{\rm{tot}} $ from the right-hand side of Eq.~\eqref{eq:Enonconserved}.
In the inner parts of galaxies, we estimate $ \dot{\phi}_N $ as roughly $ G \dot{M}(r)/r \approx (\dot{M}(r) / M(r)) \cdot \phi_N \approx (\dot{\rho}_{\rm{tot}}/\rho_{\rm{tot}}) \cdot \phi_N $, where $ M(r) \approx V \cdot \rho_{\rm{tot}} $ is the total mass inside a sphere with radius $ r $ and volume $ V $.
Since $ \phi_N \ll 1 $, we expect $ \dot{\phi}_N $ to be much smaller than $ \dot{\rho}_{\rm{tot}}/\rho_{\rm{tot}} $.
Therefore, in the inner parts of galaxies,
\begin{align}
 \label{eq:tQlocal}
 \left| \frac{\dot{\rho}_{\rm{tot}}}{\rho_{\rm{tot}}} \right| \approx \lambda \, m \, \frac{\rho_b}{\rho_{\rm{tot}}} = t_{\rm{loc}}^{-1} \approx \frac{1}{10^8\,\rm{yr}}\, \frac{\rho_b}{\rho_{\rm{tot}}} \,,
\end{align}
where we defined $ t_{\rm{loc}}^{-1} \equiv \lambda \, m \, (\rho_b/\rho_{\rm{tot}}) $ and used $ \mu \approx m $.
The timescale $ t_{\rm{loc}} $ is a local analogue of $ t_Q $.
It is the timescale on which {\sc SFDM}'s approximate equilibrium with approximately conserved $ Q $ fails to be time-independent at each point in space.
That is, locally, {\sc SFDM}'s approximate equilibrium can be valid only on timescales shorter than $ t_{\rm{loc}} $.
Thus, the time derivatives needed to satisfy the continuity equation $ \nabla_\alpha T^{\alpha 0} = 0 $ are not necessarily of order $ 1/t_Q $.
Still, both $ t_Q^{-1} $ and $ t_{\rm{loc}}^{-1} $ are proportional to $ \lambda $ since both are a consequence of $ Q $ not being conserved.

\section{An alternative model}
\label{sec:alternative}

So far we assumed $ t_Q $ to be much larger than the timescales of interest such that we could assume $ Q $ to be approximately conserved.
We will now drop this assumption and allow $ t_Q $ to take any value.
In this case, it is not generally justified to introduce a chemical potential associated with $ Q $.
As discussed above, this may be problematic for {\sc SFDM} since the chemical potential is important for {\sc SFDM}'s phenomenology of galactic rotation curves.
Here, we propose a model which avoids this problem but retains {\sc SFDM}'s phenomenology of galactic rotation curves.
We will now introduce and discuss this model.

Our model uses two complex scalar fields instead of just one,
\begin{align}
 \phi_1 = \frac{\rho_1}{\sqrt{2}} \exp(-i \theta_1)\,, \quad \phi_2 = \frac{\rho_2}{\sqrt{2}} \exp(-i \theta_2) \,,
\end{align}
and has the Lagrangian
\begin{align}
 \label{eq:Lalt}
 \mathcal{L}_{\textrm{alt}} &= \frac12 \left[ f(K_1, \rho_1) + f(K_2, \rho_2) - \lambda \, \left(\theta_1 + \theta_2\right) \, \rho_b \right] \,, \\
                 K_{\rm{i}} &= \frac12 \nabla_\alpha \rho_{\rm{i}} \, \nabla^\alpha \rho_{\rm{i}} + \frac12 \rho_{\rm{i}}^2 \,  \nabla_\alpha \theta_{\rm{i}} \, \nabla^\alpha \theta_{\rm{i}} \,,\quad \rm{i}=1,2 \,.
\end{align}
That is, we take two copies of the Lagrangian $ \mathcal{L} $ from Eq.~\eqref{eq:L} but with a global prefactor of $ 1/2 $.
This Lagrangian has an exchange symmetry $ \phi_1 \leftrightarrow \phi_2 $.
For convenience, we introduce $ \rho_+ $, $ \rho_- $, $ \theta_+ $, and $ \theta_- $ as
\begin{subequations}
\begin{align}
 \rho_1 &\equiv \rho_+ + \rho_- \,, &\theta_1 &\equiv \theta_+ + \theta_- \,,\\
 \rho_2 &\equiv \rho_+ - \rho_- \,, &\theta_2 &\equiv \theta_+ - \theta_- \,.
\end{align}
\end{subequations}
In terms of these variables, $ \mathcal{L}_{\textrm{alt}} $ has an exact shift symmetry $ \theta_- \to \theta_- + \textrm{const.} $ for $ \theta_- $, but not for $ \theta_+ $ due to the baryon coupling.
The conserved current associated with the $ \theta_- $ shift symmetry is
 \begin{align}
 \label{eq:j-}
 j_-^\alpha = \frac{\partial \mathcal{L}_{\textrm{alt}}}{\partial (\nabla_\alpha \theta_-)} = \frac12 \left[ \frac{\partial f_1}{\partial K_1} \, \rho_1^2 \, \nabla^\alpha \theta_1 - \frac{\partial f_2}{\partial K_2} \, \rho_2^2 \, \nabla^\alpha \theta_2  \right] \,,
\end{align}
with the short-hand notation $ f_{\rm{i}} \equiv f(K_{\rm{i}}, \rho_{\rm{i}}) $ for $ \rm{i} = 1, 2 $.
A similar Lagrangian was previously proposed in Ref.~\cite{Ferreira2019}.
However, our treatment is quite different in that we investigate the role of the chemical potential for the equilibrium superfluid.
Indeed, Ref.~\cite{Ferreira2019} is based on a non-relativistic limit of the form $ \theta \to m \cdot t + \theta $ with $ |\dot{\theta}|  \ll m $ which is not suitable for the equilibrium superfluid, as discussed in Sec.~\ref{sec:nonrel}.

Consider now field configurations that are symmetric under $ \phi_1 \leftrightarrow \phi_2 $, i.e. consider $ \phi_1 = \phi_2 $ or, equivalently, $ \rho_- = 0 $ and $ \theta_- = 0 $.
In this case, our model exactly recovers the equations of motion of the usual {\sc SFDM} Lagrangian from Eq.~\eqref{eq:L}.
That is, the two equations for $ \rho_1 $ and $ \rho_2 $ are equivalent and reduce to a single equation for $ \rho_+ \equiv \rho $.
This equation exactly matches the equation of motion for $ \rho $ from the usual {\sc SFDM} Lagrangian from Eq.~\eqref{eq:L}.
Similarly, the two equations for $ \theta_1 $ and $ \theta_2 $ reduce to a single equation for $ \theta_+ \equiv \theta $ which is the same as the usual {\sc SFDM} equation for $ \theta $.
Further, $ \phi_1 = \phi_2 $ implies that the {\sc EMT} $ T_{\rm{alt}, \alpha \beta} $ of our model is exactly the same as the  usual {\sc SFDM} {\sc EMT} $ T_{\alpha \beta} $.
These results are derived in Appendix~\ref{sec:alteomemt}.
The global factor of $ 1/2 $ in $ \mathcal{L}_{\textrm{alt}} $ is inconsequential for the equations of motion, but it is needed to recover {\sc SFDM}'s {\sc EMT} and {\sc SFDM}'s {\sc MOND}-like force on the baryons.

In a general nonequilibrium situation, there is no reason to expect a completely symmetric field configuration $ \phi_1 = \phi_2 $.
Therefore, we do not expect our model to exactly reproduce {\sc SFDM} in such situations.
In equilibrium, however, we expect the exchange symmetry $ \phi_1 \leftrightarrow \phi_2 $ to be respected.
Naively, this is because there is no reason for $ \phi_1 $ and $ \phi_2 $ to differ in equilibrium.
But this is not entirely correct.
In equilibrium, our model requires a chemical potential $ \mu_- $ due to the exact shift symmetry of $ \theta_- $.
As discussed in Sec.~\ref{sec:chempot}, this implies a shift of the time derivatives of $ \theta_- $,
\begin{align}
 \dot{\theta}_- \to \dot{\theta}_- + \mu_- \,.
\end{align}
In terms of $ \dot{\theta}_1 $ and $ \dot{\theta}_2 $,
\begin{subequations}
\begin{align}
 \dot{\theta}_1 \to \dot{\theta}_1 + \mu_- \,, \\
 \dot{\theta}_2 \to \dot{\theta}_2 - \mu_- \,.
\end{align}
\end{subequations}
We see that the chemical potential $ \mu_- $ induces an asymmetry between $ \phi_1 $ and $ \phi_2 $ since $ \dot{\theta}_1 $ and $ \dot{\theta}_2 $ are shifted with opposite signs.
However, it turns out that this sign of $ \mu_- $ is not important in equilibrium.
For example, the grand canonical partition function satisfies $ Z_{\textrm{alt}}(\mu_-) = Z_{\textrm{alt}}(-\mu_-) $ as discussed in Appendix~\ref{sec:altT}.
More concretely, consider the zero-temperature superfluid.
We assume that all zero-temperature quantities can be calculated using the equations of motion of the effective Lagrangian $ \mathcal{L}_{\textrm{alt}}(\dot{\theta}_- + \mu_-, \dots) $.
Appendix~\ref{sec:alteomemt} then shows that the sign of $ \mu_- $ does not enter these equilibrium equations such that they are symmetric under $ \phi_1 \leftrightarrow \phi_2 $.
This follows for a static background metric by assuming that all fields are time independent in equilibrium.
Therefore, we expect that the exchange symmetry $ \phi_1 \leftrightarrow \phi_2 $ is respected in equilibrium.
That is, we assume $ \phi_1 = \phi_2 $ for the zero-temperature superfluid.
Or equivalently, $ \rho_- = 0 $ and $ \theta_- = 0 $.
Here, $ \theta_- = 0 $ means that we first do the shift $ \dot{\theta}_- \to \dot{\theta}_- + \mu_- $ and then apply $ \theta_- = 0 $.

The resulting zero-temperature equations exactly recover {\sc SFDM}'s zero-temperature equations of motion, including a chemical potential $ \mu_- \equiv \mu $.
Likewise, our model recovers {\sc SFDM}'s zero-temperature energy density.
This is shown in Appendix~\ref{sec:alteomemt}.
Therefore, our model recovers {\sc SFDM}'s phenomenology of galactic rotation curves in the zero-temperature superfluid cores of galaxies.
Importantly, it does so without relying on a condition like $ t_Q \gg t_{\textrm{dyn}} $.
This is in contrast to the usual {\sc SFDM} Lagrangian from Eq.~\eqref{eq:L} which allows introducing a chemical potential only on timescales much shorter than $ t_Q $.
This also implies that our model's equilibrium {\sc EMT} $ T_{\textrm{alt},\alpha \beta} $ exactly satisfies the continuity equation $ \nabla_\alpha T^{\alpha 0}_{\textrm{alt}} = 0 $, while this is true for {\sc SFDM} only up to time derivatives of order $ 1/t_Q $ (see Sec.~\ref{sec:emt}).
Concretely, we have $ T_{\textrm{alt}, 0j} = 0 $ for our model, but $ T_{0j} \neq 0 $ for {\sc SFDM} for $ \rm{j} = 1, 2, 3 $, as discussed in Appendix~\ref{sec:alteomemt}.
This is not an important difference, however, since these $ 0j $ components are usually neglected in actual calculations \cite{Berezhiani2015, Berezhiani2018}.

Above, we assumed that our model's equilibrium is constrained by the shift symmetry of $ \theta_- $ but not by the broken shift symmetry of $ \theta_+ $.
Reaching this equilibrium requires an interaction that respects the shift symmetry of $ \theta_- $ but not that of $ \theta_+ $.
In $ \mathcal{L}_{\textrm{alt}} $, this leaves only the baryon coupling which is $ \lambda $-suppressed.
As a result, reaching equilibrium may take a very long time.
Possibly, it will never be reached on galactic scales.
To remedy this, we may introduce an additional coupling that breaks the $ \theta_+ $ shift symmetry, e.g.
\begin{align}
 \lambda_m \, \rho_-^2 \, \theta_+^2 \,,
\end{align}
with constant $ \lambda_m $.
This term has the same symmetries as $ \mathcal{L}_{\textrm{alt}} $ so that we still have an equilibrium with a chemical potential $ \mu_- $ that respects the $ \phi_1 \leftrightarrow \phi_2 $ symmetry.
In particular, $ \rho_- $ and $ \theta_- $ still vanish in equilibrium.
As a result, this term allows reaching equilibrium, but it does not affect the zero-temperature equilibrium itself due to the factor $ \rho_-^2 $.
We leave a detailed study of this coupling for future work.

As discussed in Ref.~\cite{Berezhiani2015}, a realistic {\sc SFDM} model may require finite-temperature corrections.
Here, our model differs from the usual {\sc SFDM} Lagrangian from Eq.~\eqref{eq:L}.
For $ \lambda_m = 0 $, the main reason is that our model has more degrees of freedom.
As a very simple example, take $ f(K, \rho) = K - \frac12 m^2 \rho^2 $ and $ \rho_b = 0 $.
Then, $ \mathcal{L} $ is the Lagrangian of a single complex scalar field and we have $ T_{00} \neq 0 $ due to thermal fluctuations.
Further, $ \phi_1 $ and $ \phi_2 $ can be rescaled such that $ \mathcal{L}_{\textrm{alt}} $ is the Lagrangian of two such fields.
This implies $ T_{\textrm{alt},00} = 2 T_{00} \neq T_{00} $.
For $ \lambda_m \neq 0 $, differences in finite-temperature corrections may be more interesting due to direct interactions between $ \phi_1 $ and $ \phi_2 $.
In any case, as a consequence of the dissipation-fluctuation theorem, our model also differs from {\sc SFDM} regarding perturbations on top of the equilibrium superfluid \cite{Livi2017}.
This may be important for stability analyses \cite{Berezhiani2015}.
An explicit calculation of these perturbations or of the finite-temperature corrections is beyond the scope of this paper, however.
For the usual {\sc SFDM} Lagrangian from Eq.~\eqref{eq:L}, calculations at finite temperature were done in Ref.~\cite{Sharma2019}, but so far only for a simplified model without a {\sc MOND} limit.

\section{Conclusion}
\label{sec:conclusion}

In this paper, we have clarified several issues regarding {\sc SFDM}'s chemical potential.
First, we have shown that introducing a chemical potential is justified only if we are interested in timescales much shorter than $ t_Q \approx (M_{\textrm{DM}}/M_b) \cdot 10^8\,\textrm{yr} $ (or its local analogue $ t_{\rm{loc}} \approx (\rho_{\rm{tot}}/\rho_b) \cdot 10^8\,\textrm{yr} $).
This is because the nonconservation of the $ U(1) $ charge can be neglected only on these timescales.
We have then shown that correctly introducing the chemical potential does not introduce an explicit time dependence in the symmetry-breaking baryon-phonon coupling.
This is because the chemical potential is correctly introduced by shifting only time derivatives of $ \theta $, but not occurrences of $ \theta $ without a time derivative.
As a consequence, we could distinguish two different non-relativistic limits which are appropriate in different physical situations.
One is appropriate for particle-like solutions in vacuum, the other for the superfluid cores of galaxies.
Finally, we have proposed a model that recovers {\sc SFDM}'s zero-temperature equations including a chemical potential even if the timescales of interest are not much smaller than $ t_Q $.
This is possible by introducing two complex scalar fields $ \phi_1 $ and $ \phi_2 $ with an exchange symmetry $ \phi_1 \leftrightarrow \phi_2 $ in such a way that one linear combination of their phases has an exact shift symmetry.

Our results help to better understand possible microscopic realizations of {\sc SFDM}, especially with respect to issues regarding the chemical potential and the nonconservation of charge.
This improved understanding as well as the alternative model from Sec.~\ref{sec:alternative} may prove helpful in {\sc SFDM} model building.

\section*{Acknowledgements}
\label{sec:acknowledgements}

The author thanks Sabine Hossenfelder for valuable discussions and for reading the manuscript.

\begin{appendices}

\section{$ \mathcal{L}_{\rm{alt}} $ equations of motion and {\sc EMT}}
\label{sec:alteomemt}

Here, we will discuss the equations of motion and the {\sc EMT} of our model from Sec.~\ref{sec:alternative}.

We will first show that this model recovers the usual {\sc SFDM} equations of motion for $ \phi_1 = \phi_2 $.
This works because our model is symmetric under $ \phi_1 \leftrightarrow \phi_2 $.
More explicitly, consider the equations of motion for $ \rho_1 $ and $ \rho_2 $.
They are
\begin{align}
 \label{eq:eomrho12}
 \nabla_\alpha \left( \frac{\partial f_{\rm{i}}}{\partial K_{\rm{i}}} \, \nabla^\alpha \rho_{\rm{i}} \right) - \frac{\partial f_{\rm{i}}}{\partial K_{\rm{i}}} \rho_{\rm{i}} \left(\nabla_\alpha \theta_{\rm{i}} \, \nabla^\alpha \theta_{\rm{i}} \right) - \frac{\partial f_{\rm{i}}}{\partial \rho_{\rm{i}}} = 0 \,, \quad \rm{i} = 1,2 \,.
\end{align}
For $ \phi_1 = \phi_2 $, we have $ \rho_1 = \rho_2 = \rho_+ \equiv \rho $ and $ \theta_1 = \theta_2 = \theta_+ \equiv \theta $.
As a result, the two equations in Eq.~\eqref{eq:eomrho12} reduce to a single equation for $ \rho_+ $ which is the same equation as the equation for $ \rho $ in the case of the usual {\sc SFDM} Lagrangian from Eq.~\eqref{eq:L}.
Similarly, the equations of motion for $ \theta_1 $ and $ \theta_2 $ are
\begin{align}
 \label{eq:eomtheta12}
 \nabla_\alpha \left( \frac{\partial f_{\rm{i}}}{\partial K_{\rm{i}}} \, \rho_{\rm{i}}^2 \, \nabla^\alpha \theta_{\rm{i}} \right) = - \lambda \rho_b \,, \quad \rm{i} = 1,2 \,.
\end{align}
Due to $ \phi_1 = \phi_2 $, these reduce to a single equation for $ \theta_+ $ which is the same equation as the usual {\sc SFDM} equation for $ \theta $.
Further, the {\sc EMT} of our model is
\begin{align}
 \label{eq:emt12}
 T_{\textrm{alt},\alpha \beta} = \frac12 \sum_{\rm{i} = 1}^{2} \left[ \frac{\partial f_{\rm{i}}}{\partial K} \left( \nabla_\alpha \rho_{\rm{i}} \, \nabla_\beta \rho_{\rm{i}} + \rho_{\rm{i}}^2 \, \nabla_\alpha \theta_{\rm{i}} \, \nabla_\beta \theta_{\rm{i}} \right) - g_{\alpha \beta} (f_{\rm{i}} - \lambda \, \theta_{\rm{i}} \, \rho _b ) \right] \,.
\end{align}
For $ \phi_1 = \phi_2 $, this {\sc EMT} agrees exactly with that of {\sc SFDM} from Eq.~\eqref{eq:emt} since the two terms for $ \rm{i} = 1,2 $ in Eq.~\eqref{eq:emt12} are identical such that they cancel the prefactor of $ 1/2 $.

Next, consider the zero-temperature equilibrium equations of motion.
We assume the same static background metric as in Sec.~\ref{sec:chempot}.
We introduce a chemical potential $ \mu_- $ associated with the exact shift symmetry of $ \theta_- $ by shifting $ \dot{\theta}_- \to \dot{\theta}_- + \mu_-  $.
This shifts $ \theta_1 $ and $ \theta_2 $ with opposite sign, as discussed in Sec.~\ref{sec:alternative}.
Despite this, the equations of motion of $ \phi_1 $ and $ \phi_2 $ are symmetric under $ \phi_1 \leftrightarrow \phi_2 $, if all fields are time independent.
To see this, consider the $ \mu_- $-dependence of $ K_1 $ and $ K_2 $,
\begin{subequations}
\begin{align}
 K_1 = \frac12 \rho_1^2 \, g^{00} \left( \dot{\theta}_1 + \mu_- \right)^2 + \dots \,, \\
 K_2 = \frac12 \rho_2^2 \, g^{00} \left( \dot{\theta}_2 - \mu_- \right)^2 + \dots \,,
\end{align}
\end{subequations}
We see that $ \mu_- $ enters $ K_1 $ and $ K_2 $ with opposite sign.
But for time-independent fields, this sign cancels due to the square.
In this case, $ K_1 $ and $ K_2 $ are symmetric under $ \phi_1 \leftrightarrow \phi_2 $.
It follows that the equations of motion for $ \rho_1 $ and $ \rho_2 $ are also symmetric under $ \phi_1 \leftrightarrow \phi_2 $.
This is because these equations depend on $ \mu_- $ only through $ K_1 $ and $ K_2 $.
In the equations of motion of $ \theta_1 $ and $ \theta_2 $, there is an additional dependence on $ \mu_- $,
\begin{subequations}
\begin{align}
 \frac{1}{\sqrt{-g}} \, \partial_\alpha \left( \sqrt{-g} \, \frac{\partial f_1}{\partial K_1} \, \rho_1^2 \left[\nabla^\alpha \theta_1 + g^{\alpha 0} \mu_- \right] \right) = -\lambda \rho_b \,, \\
 \frac{1}{\sqrt{-g}} \, \partial_\alpha \left( \sqrt{-g} \, \frac{\partial f_2}{\partial K_2} \, \rho_2^2 \left[\nabla^\alpha \theta_2 - g^{\alpha 0} \mu_- \right] \right) = -\lambda \rho_b \,.
\end{align}
\end{subequations}
However, this additional $ \mu_-$-dependence is multiplied by a time derivative which vanishes for time-independent fields.
Therefore, also the equations of motion of $ \theta_1 $ and $ \theta_2 $ are symmetric under $ \phi_1 \leftrightarrow \phi_2 $.
It follows that the zero-temperature equilibrium equations of motion of {\sc SFDM} including a chemical potential $ \mu_- \equiv \mu $ are recovered for $ \phi_1 = \phi_2 $.

However, our model does not fully recover {\sc SFDM}'s equilibrium {\sc EMT}.
This is because the asymmetry induced by $ \mu_- $ is significant in the {\sc EMT}.
In particular, consider the {\sc EMT} of our model from Eq.~\eqref{eq:emt12}, but with a shift $ \dot{\theta}_- \to \dot{\theta}- + \mu_ -$.
The sign of $ \mu_- $ is not important in the $ \mu_- $-dependence of $ K_1 $ and $ K_2 $ as discussed above.
Therefore, the $ -(1/2) g_{\alpha \beta}(f_i - \lambda \, \theta_{\rm{i}} \, \rho_b) $ terms are symmetric under $ \phi_1 \leftrightarrow \phi_2 $ and reproduce the corresponding term of {\sc SFDM}'s {\sc EMT} for $ \phi_1 = \phi_2 $.
This leaves the terms
\begin{align}
 \frac12 \left( \frac{\partial f_1}{\partial K_1} \rho_1^2 \, (\partial_\alpha \theta_1 + \delta_\alpha^0 \mu_-) \, (\partial_\beta \theta_1 + \delta_\beta^0 \mu_-) + \frac{\partial f_2}{\partial K_2} \rho_2^2 \, (\partial_\alpha \theta_2 - \delta_\alpha^0 \mu_-) \, (\partial_\beta \theta_2 - \delta_\beta^0 \mu_-) \right) \,.
\end{align}
If either both or none of $ \alpha $ and $ \beta $ are zero, the sign of $ \mu_- $ drops out for time-independent fields.
Thus, for $ \phi_1 = \phi_2 $, our model recovers the corresponding components $ T_{\alpha \beta} $ of {\sc SFDM}'s equilibrium {\sc EMT}.
This leaves the case $ \alpha = j $, $ \beta = 0 $ for $ j = 1, 2, 3 $.
For time-independent fields and for $ \phi_1 = \phi_2 $, this gives
\begin{align}
 \frac12 \frac{\partial f}{\partial K} \rho^2 \left( (\partial_j \theta) \, \mu_- + (\partial_j \theta) \, (-\mu_-) \right) = 0 \,.
\end{align}
where we have identified $ \rho_+ \equiv \rho $, $ \theta_+ \equiv \theta $, $ K_i \equiv K $, and $ f_i \equiv f $ for $ \rm{i} = 1, 2 $.
We see that the two terms for $ \rm{i} = 1, 2 $ cancel each other due to the asymmetry induced by $ \mu_- $.
Thus, we have $ T^{0j}_{\textrm{alt}} = 0 $, while {\sc SFDM}'s usual equilibrium {\sc EMT} has $ T^{0j} \neq 0 $, see Eq.~\eqref{eq:T0j}.

\section{Chemical potential dependence of $ Z_{\textrm{alt}} $}
\label{sec:altT}

In this appendix, we will show that the grand canonical partition function $ Z_{\textrm{alt}} $ of $ \mathcal{L}_{\textrm{alt}} $ satisfies $ Z_{\textrm{alt}}(\mu_-) = Z_{\textrm{alt}}(-\mu_-) $.
To obtain this result, we employ a spurion analysis.
That is, we first note that $ \mathcal{L}_{\textrm{alt}} $ has a symmetry $ (\theta_1, \theta_2, \lambda) \to (-\theta_1, -\theta_2, -\lambda) $.
This is a generalized charge conjugation symmetry that changes the sign of both $ Q_+ $ and $ Q_- $.
It implies that the sign of $ \lambda $ is unphysical.
Next, we promote $ \lambda $ from a parameter to a field that spontaneously breaks this symmetry.
This does not change the phenomenology of our model if we make $ \lambda $ very heavy.
Concretely, this means we introduce a potential $ \alpha \, ( \lambda^2 - \lambda_0^2)^2 $ for $ \lambda $ with very large $ \alpha $.
Then, $ \lambda $ obtains an equilibrium value $ |\lambda| = |\lambda_0| $.
This is because all other terms can be neglected for large enough $ \alpha $.
In particular, the baryon-phonon coupling $ - (1/2) \lambda \, (\theta_1 + \theta_2) \, \rho_b $ does not influence the $ \lambda $ equilibrium value if $ \alpha $ is large enough.
We will exploit this to show $ Z_{\textrm{alt}}(\mu_-) = Z_{\textrm{alt}}(-\mu_-) $.
We start with the grand canonical partition function $ Z_{\textrm{alt}}(\mu_-) $ including the field $ \lambda $,
\begin{multline}
 \label{eq:Zaltmu}
 Z_{\textrm{alt}}(\mu_-) = \int \mathcal{D}\pi_1 \, \mathcal{D}\theta_1 \, \mathcal{D}\pi_2 \, \mathcal{D}\theta_2 \, \mathcal{D}\pi_\lambda \, \mathcal{D}\lambda \exp\left[ \int d(it) \int d^3\vec{x} \right. \\
                                \left. \times \left(\pi_1 \,\dot{\theta}_1 + \pi_2 \, \dot{\theta}_2 + \pi_\lambda \dot{\lambda} - \sqrt{-g} \, \mathcal{H}_{\textrm{alt}}(\pi_1, \theta_1, \pi_2, \theta_2, \pi_\lambda, \lambda) + \mu_- (\pi_1 - \pi_2) \right) \right] \,.
\end{multline}
Here, $ \mathcal{H}_{\textrm{alt}} $ is the Hamiltonian density derived form $ \mathcal{L}_{\textrm{alt}} $ including the field $ \lambda $.
Then, we perform a change of variables that corresponds to the generalized charge conjugation mentioned above.
That is, we switch the sign of all fields in the path integral in Eq.~\eqref{eq:Zaltmu},
\begin{align}
 \pi_1 \to -\pi_1 \,, \quad \theta_1 \to -\theta_1  \,,\quad \pi_2 \to -\pi_2  \,,\quad \theta_2 \to -\theta_2  \,,\quad \pi_\lambda \to -\pi_\lambda  \,,\quad \lambda \to -\lambda \,.
\end{align}
Due to the symmetry of $ \mathcal{L}_{\textrm{alt}} $, we have
\begin{align}
 \mathcal{H}_{\textrm{alt}}(\pi_1, \theta_1, \pi_2, \theta_2, \pi_\lambda, \lambda) = \mathcal{H}_{\textrm{alt}}(-\pi_1, -\theta_1, -\pi_2, -\theta_2, -\pi_\lambda, -\lambda) \,.
\end{align}
Therefore, the effect of this change of variables is simply to change the sign of the $ \mu_- (\pi_1 - \pi_2) $ term in the exponent of Eq.~\eqref{eq:Zaltmu}.
Otherwise, Eq.~\eqref{eq:Zaltmu} is unaffected.
Thus,
\begin{align}
 Z_{\textrm{alt}}(\mu_-) = Z_{\textrm{alt}}(-\mu_-) \,.
\end{align}
With the same argument, we can also show that the equilibrium ensemble averages of all observables $ O $ with
\begin{align}
 O(\pi_1, \theta_1, \pi_2, \theta_2, \pi_\lambda, \lambda) = O(-\pi_1, -\theta_1, -\pi_2, -\theta_2, -\pi_\lambda, -\lambda)
\end{align}
do not depend on the sign of $ \mu_- $.
These are the observables that are invariant under the generalized charge symmetry mentioned above.

Alternatively, $ Z_{\textrm{alt}}(\mu_-) = Z_{\textrm{alt}}(-\mu_-) $ can be obtained by using the $ \phi_1 \leftrightarrow \phi_2 $ symmetry of $ \mathcal{L}_{\textrm{alt}} $.
We chose the spurion analysis since it also works for the approximate equilibrium of {\sc SFDM} on timescales much shorter than $ t_Q $, i.e. we have $ Z(\mu) = Z(-\mu) $.

\end{appendices}

\printbibliography[heading=bibintoc]

\end{document}